\documentclass[aps,prl,twocolumn,superscriptaddress]{revtex4}

\usepackage{dcolumn}
\usepackage{bm}

\usepackage{epsfig}
\usepackage{graphicx}
\usepackage{color}
\usepackage{amsmath,amssymb}

\newcommand{\beq}[1]{
\begin{equation}
\label{e#1} }

\newcommand{\eeq}{
\end{equation}
}

\begin{document}

\title{Electrical excitation and detection of magnetic dynamics with impedance matching}
\author{D.~Fang}
\affiliation{Hitachi Cambridge Laboratory, Cambridge CB3 0HE, United Kingdom}
\author{T.~Skinner}
\affiliation{Cavendish Laboratory, University of Cambridge, CB3 0HE, United Kingdom}
\author{H.~Kurebayashi}
\affiliation{Cavendish Laboratory, University of Cambridge, CB3 0HE, United Kingdom}
\author{R.~P.~Campion}
\affiliation{School of Physics and
Astronomy, University of Nottingham, Nottingham NG7 2RD, United Kingdom}
\author{B.~L.~Gallagher}
\affiliation{School of Physics and
Astronomy, University of Nottingham, Nottingham NG7 2RD, United Kingdom}
\author{A.~J.~Ferguson}
\email{ajf1006@cam.ac.uk}
\affiliation{Cavendish Laboratory, University of Cambridge, CB3 0HE, United Kingdom}

\date{\today}

%

\begin{abstract}
Motivated by the prospects of increased measurement bandwidth, improved signal to noise ratio and access to the full complex magnetic susceptibility we develop a technique to extract microwave voltages from our high resistance ($\sim$ 10 k$\Omega$) (Ga,Mn)As microbars. We drive magnetization precession with microwave frequency current, using a mechanism that relies on the spin orbit interaction. A capacitively coupled $\lambda/2$ microstrip resonator is employed as an impedance matching network, enabling us to measure the microwave voltage generated during magnetisation precession.
\end{abstract}
\maketitle

\begin{figure}
  \includegraphics[width=6cm]{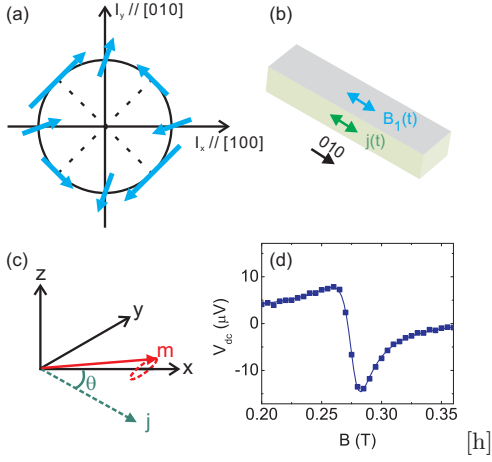} [h]\\
  \caption{(Color online) (a), Schematic showing the current induced effective magnetic field for compressively strained (Ga,Mn)As. The vectors represent this field for electrical currents in different crystal directions. The symmetry of the effective magnetic fields resembles the Dresselhaus and Rashba spin orbit interactions observed in two-dimensional III-V semiconductors. (b) In this work a bar in the [010] direction is studied. Passing a current in this direction results in an in plane effective magnetic directed (mostly) along the bar. Our two-terminal sample has a width of 4 $\mu$m, is 40 $\mu$m long and is etched from 25 nm thick (Ga$_{0.94}$,Mn$_{0.06}$)As layer epitaxially grown on a GaAs substrate. The sample resistance is 11.2 k$\Omega$ at 300 K and 9.1 k$\Omega$ at 30 K. In order to maximise our signal we mounted the sample at 45 degrees to the B-field, maximising $\sin{\theta} \sin{2\theta}$ and therefore $V_{dc}$ and $V_\omega$ (see equations 1 and 2). (c) The angle $\theta$ is the in-plane angle between the magnetisation and current. Magnetic field is along the x-direction. Magnetization precesses ecliptically around the external field causing $\theta$ to vary. (d) Ferromagnetic resonance observed for the [010] sample via the rectification mechanism, yielding an easily measured dc voltage. Since the microwaves are pulse modulated and the resulting dc voltage is lock-in detected, the output of the lock-in amplifiers are multplied by $2\sqrt{2}$ to give the correct amplitude of $V_{dc}$. For this measurement the frequency is 6.804 GHz, the power at the sample is determined to be -5 dBm and the temperature is 30 K. Less than 1 K temperature rise in the sample is observed on application of the microwave signal.}
\end{figure}

Magneto-resistance effects displayed by ferromagnets and ferromagnetic devices enable the observation of precessing magnetization. A direct-current passed through the sample results in a oscillating voltage due to the oscillating magneto-resistance \cite{Tulapurkar:2005_a}. Alternatively if a microwave frequency current is passed through the sample, the sample itself can be used as a rectifier: the oscillating magneto-resistance multiplied by the oscillating current yields a time-independent voltage \cite{Juretschke:1960_a,Costache:2006_a, Meking:2007_a,Sankey:2006_a,Yamaguchi:2008_a}. It is straightforward to extract the microwave voltage in the case of samples with a resistance of about 50~$\Omega$, however significantly higher (or lower) resistance samples suffer from an impedance mismatch problem. In this Letter we demonstrate a simple impedance matching technique that can be used to extract this microwave voltage from high-resistance samples. We use a $\lambda/2$ microstrip resonator to impedance match a $\sim$ 10~k$\Omega$ sample towards 50~$\Omega$ at frequency of 7 GHz. We are motivated to access the microwave frequency voltage because of the advantages it confers over the dc rectification signal, in the following we list a few. The measurement bandwidth is determined by the Q-factor of the resonator, giving 350 MHz bandwidth rather than the $\sim$ kHz offered by rectification; the use of low noise microwave amplifiers enables an improved signal to noise ratio; the microwave voltage allows access to the real and imaginary parts of the complex susceptibility; and the microwave voltage is ideally suited to studying the bias dependence of spin transfer torques \cite{Xue_2012}. Finally, we anticipate that the study of the microwave voltage in thin film ferromagnetic layers will lead to the discovery of new phenomena in magnetisation dynamics.

\begin{figure}
  \includegraphics[width=6cm]{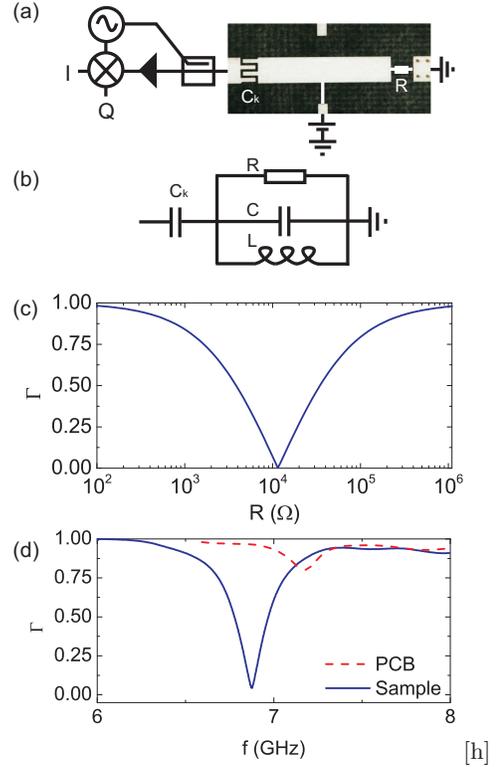}[h]\\
  \caption{(Color online) (a), Photograph of the resonator PCB included in a schematic of the microwave reflectometry set-up. The ground plane at the end of the resonator (coupled to ground with 4 plated via holes), and the tabs for making the bias tee are visible. In the coupling capacitor dimensions of the fingers are $400~\mu$m$ \times 1.5$~mm and they are separated from each other by $200~\mu$m. The microstrip is 2.5 mm wide and 12.5~mm long and fabricated on a 0.787 mm thick substrate, Rogers 5880 ($\epsilon_r$=2.2, tan $\delta$ = 0.0009 at 10 GHz). The PCB is plated with immersion silver. The reflectometry setup includes a 10 dB directional coupler (PE2204-10), a 25 dB amplifier (PE1522) for the reflected signal and an IQ mixer (IQ-4509-MXP). The mixer output is further amplified by 20 dB with a $50~\Omega$ input impedance amplifier (SR445A). (b) Equivalent circuit of the transmission line resonator, coupling capacitor and sample. (c) Expected reflection coefficient plotted as a function of frequency using the impedance given in eqn. 4. A coupling capacitance of 30 fF was used together with the expected inductance (L=0.73 nH) and capacitance (C=0.71 pF) of our resonator. (d) Room temperature reflectometry measurements of the resonator with ($Q=20$) and without ($Q=37$) the sample attached, cable loss has been subtracted.}
\end{figure}

We use normal-metal $\lambda/2$ resonators but note that high quality factor superconducting resonators \cite{Frunzio_2005,Goppl_2008} have been widely used in radio-astronomy and condensed matter physics over the past decade. These superconducting resonators have been used for sensitive detection of x-rays\cite{Mazin_2002}, read-out of superconducting qubits \cite{Wallraff_2004} and double quantum dots\cite{Frey_2012} and are also of interest for high-sensitivity electron spin resonance \cite{Schuster_2010, Kubo_2010, Wu_2010}.

Passing an electrical current thorough the dilute magnetic semiconductor (Ga,Mn)As \cite{Jungwirth:2006_a} generates an effective magnetic field \cite{Garate:2009_a,Manchon:2009_a,Chernyshov:2009_a,Endo:2010_a,Fang:2010_a}. The origin is the combination of spin accumulation due to the spin-orbit coupled bandstructure (the inverse spin galvanic effect\cite{Edelstein:1990_a}) and the exchange coupling between the carriers and the local moments. As this is a bandstructure effect, the direction of the current induced field depends on the current direction with respect to the crystal (fig. 1(a)). We use a sample in the [010] direction producing an in-plane field nearly parallel to the current direction (fig. 1(b)). For typical samples the magnitude of the current induced field is 1 mT/$10^6$ Acm$^{-2}$. The current induced field provides a convenient way of driving ferromagnetic resonance (FMR) \cite{Fang:2010_a} and the oscillating magnetic field ($B_1e^{j\omega t}$) can be in the range 1 $\mu$T - 1 mT.

\begin{figure}
  \includegraphics[width=6.0cm]{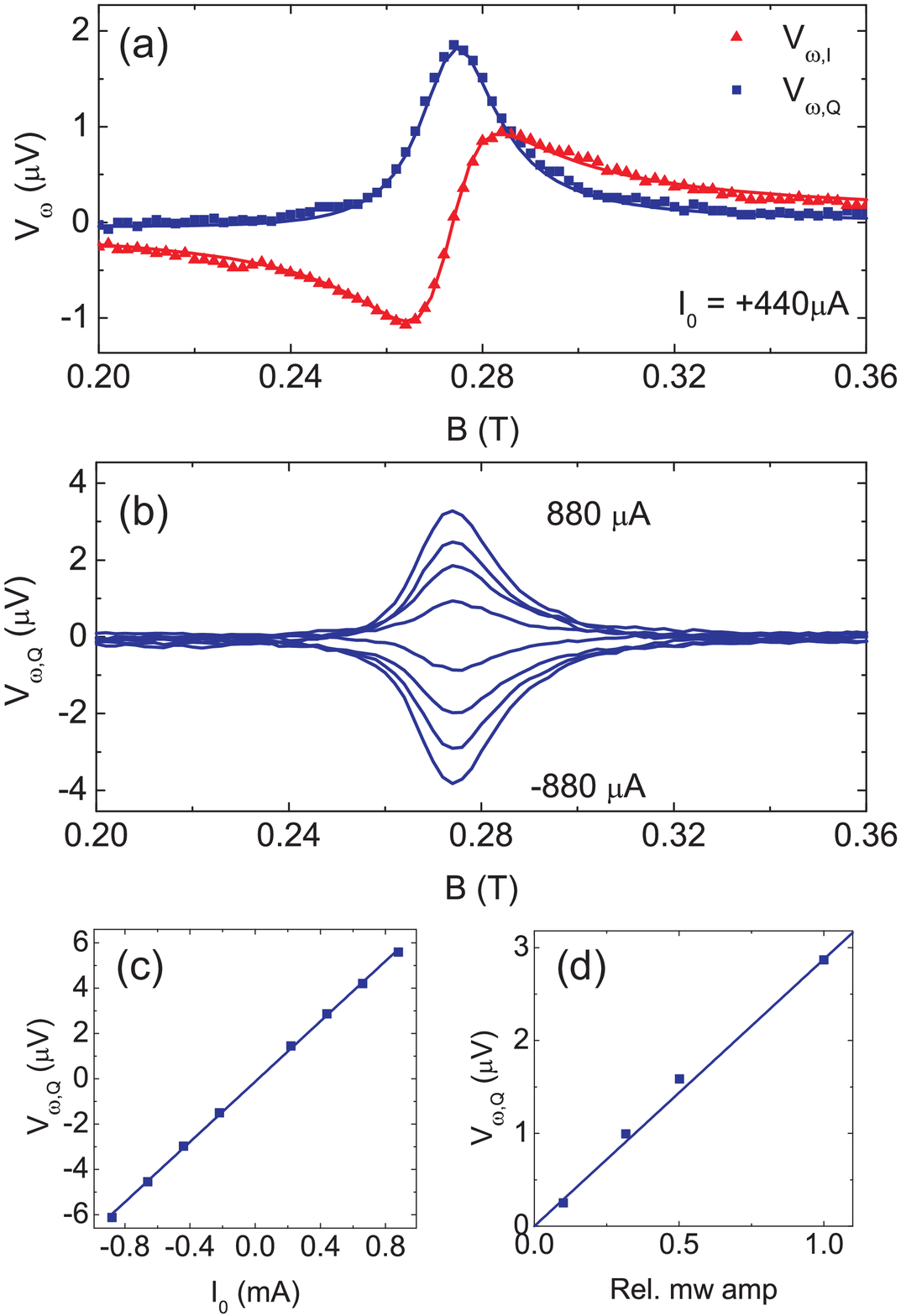}[h]\\
  \caption{(Color online) (a) I and Q outputs from the mixer showing anti-symmetric and symmetric Lorentzian forms. The sample is biased by 4 V producing a current $I_0$ of 440 $\mu A$. For all the measurements in this figure the microwave frequency is f=6.804 GHz, power at the sample is determined to be -5 dBm and sample temperature is 30 K. The lock-in amplifiers measure the RMS voltage induced by the current modulation, their output is multiplied by $2\sqrt{2}$ to give the total change in microwave signal on current modulation. Finally this is divided by the voltage gain (G=63 or 36 dB) of the system to give the components of $V_{\omega}$. At the maximum current though the system a 2 K increase of sample temperature is observed. (b) $V_{\omega,Q}$ as a function of the applied dc bias. (c) Amplitude of the symmetric Lorentzian component of $V_{\omega,Q}$ as a function of dc bias. (d) Amplitude of $V_{\omega,Q}$ as a function of microwave current $I_1$ (relative to $I_1=160$ $\mu$A) passing through the sample.}
\end{figure}

Due to the anisotropic magneto-resistance the sample resistance depends on the in-plane angle between the current and magnetisation ($\theta$) as follows: $R=R_0+\Delta R \sin{^2\theta}$ (fig. 1(c)). During precession $\theta$ varies, leading to a time-dependent change in resistance, $\delta R(t)=\frac{\Delta R m_y(t)\sin{2\theta}}{m_s}$, where $m_y(t)$ is the y-component of the magnetization. By solving the Landau-Lifshitz-Gilbert equation, $m(t)$ can be related to $B_1(t)$ by a susceptibility tensor ($m(t)=(\chi+j\chi')B_1(t)$). If $B_1e^{j\omega t}$ is along the bar, as for our [010] samples, then we can consider the following individual components of the susceptibility: $\chi_{yy}$ (an anti-symmetric Lorentzian) and  $\chi_{yy}'$ (a symmetric Lorentzian) \cite{Costache:2006_a, Meking:2007_a,Yamaguchi:2008_a,Fang:2010_a}. In the co-ordinates we use, x is along the average magnetisation direction, so the y-component of $B_1(t)$ is $-B_1(t)\sin{\theta}$. If there is a microwave frequency current $I_1 e^{j\omega t}$ through the bar, in phase with $B_1(t)$, a time independent voltage $V_{dc}$ results from Ohm's law:

\begin{equation}
V_{dc}=\frac{-\Delta RI_1 \chi_{yy}B_1\sin{\theta} \sin{2\theta}}{2m_s}
\end{equation}

This leads to the anti-symmetric Lorentzian lineshape (fig. 1(d)). A symmetric Lorentzian component is also present in the signal indicating a component of $B_1(t)$ out of plane, or a phase shift between $B_1(t)$ and $I_1(t)$. In the case of a dc current ($I_0$) through the bar a microwave voltage $V_\omega(t)$ results at the precession frequency:

\begin{equation}
V_{\omega}(t)=\frac{-\Delta RI_0(\chi_{yy}+j\chi_{yy}')B_1\sin{\theta} \sin{2\theta}e^{j\omega t}}{m_s}
\end{equation}

$V_\omega(t)$ has been studied in low-resistance spin-valve structures \cite{Tulapurkar:2005_a,Xue_2012}, and provides the most straightforward approach to measuring the bias dependence of current induced torques. Unlike $V_{dc}$ it contains information about the full complex susceptibility rather than just the real part. Some impedance matching approach must be taken to extract this signal from our samples. The voltage reflection coefficient between a 10 k$\Omega$ sample and a 50~$\Omega$ coaxial cable, is $\Gamma=\frac{R-Z_0}{R+Z_0}= 0.995$ so only 1 \% ($1-\Gamma^2$) of the incident microwave power is transmitted to the cable from our device. Of course, the same impedance mismatch problem occurs when trying to drive a microwave current through the sample. To quantify this, we measured the increase in $V_{dc}$ when impedance matching is used. We expect a 100-fold increase in $V_{dc}$ (since $V_dc\propto I_1^2$) however, due to losses in our resonator, a 48-fold increase was observed.

Now we describe our implementation of the $\lambda/2$ impedance matching network. A $Z_0$=50~$\Omega$ microstrip resonator is patterned on a low-loss printed circuit board (PCB) (fig. 2(a)). The resonator is excited through a 4-finger interdigitated capacitor and the thin film ferromagnetic sample is wire-bonded between the resonator and the ground-plane. When driven at its fundamental frequency, there is a node of electric field at the centre-point of the resonator. This enables the simple incorporation of a bias-tee \cite{Chen_2011}, a wire bond ($\approx$5 mm) is made to the centre-point and then attached to our dc circuitry. The bias-tee is observed to have negligible effect on the microwave properties of the resonator and we measure $>$ 18 dB isolation between the resonator input and the bias-tee connection. To reduce radiation losses the PCB is placed in a copper enclosure.

The impedance of the resistively-loaded resonator is described by the following expression, where l is the resonator length, $v_p=\omega/k$ is the phase velocity, R is the sample resistance and $C_k$ the coupling capacitance:

\begin{equation}
Z(\omega)=\frac{1}{j\omega C_k}+Z_0 \frac{R+jZ_0\tan(\frac{\omega l}{v_p})}{Z_0+jR\tan(\frac{\omega l}{v_p})}
\end{equation}

Simplifying equation 1, it may be seen that the $\lambda/2$ resonator is equivalent to a parallel circuit \cite{Pozar}, the sample resistance (R) in parallel with a capacitance ($C=\frac{\pi}{2 Z_0 \omega_0}$) and inductance ($L=\frac{1}{\omega_0^2 C}$) (fig. 2(b)). The resonant frequency of the unloaded circuit is given by $\omega_0^2=1/LC$. A frequency of 7 GHz gives values of C=$0.71$~pF and $L=0.73$~nH. At resonance, the impedance of the capacitatively driven parallel resonant circuit becomes real:

\begin{equation}
Z(\omega_0)\approx\frac{1}{R\omega_0^2 C_k^2}
\end{equation}

The circuit acts to invert the impedance of the resistor: $Z(\omega_0)\propto1/R$. Also notice the $1/C_k^2$ dependence, the coupling capacitance is used to define the matching resistance. Taking the expected values of L and C for our resonator and a realistic value for the coupling capacitance, $C_k=30$ fF, we show how the load resistance affects the reflection coefficient (fig. 2(c)). The load resistance is matched when $\Gamma$=0, occurring for a resistance of $\approx 10$ k$\Omega$.

The frequency response of our resonator with and without the sample attached is shown (fig. 2(d)). With the sample attached, the reflection coefficient $\Gamma_S\approx0$ at resonance indicative that the sample is close to perfectly impedance matched. With no sample attached $\Gamma_{NS}\approx0.8$ showing that conductor and dielectric losses are also contributing to, but not dominating, power loss in the resonator. These reflection coefficients help us calibrate the microwave current in the sample. Using a calibrated microwave diode we determine that the power reaching the sample is $P_{in}$=-5 dBm (320 $\mu$W). Equating the power dissipated in the sample ($(1-\Gamma_S^2-\Gamma_{NS}^2)P_{in}\approx0.36P_{in}$) to $I_1^2R/2$ we find that $I_1=160~\mu$A, giving a peak current density of $1.6\times10^5$ Acm$^{-2}$.

In order to detect $V_{\omega}(t)$ we perform microwave reflectometry. We drive the resonator close to its resonant frequency, using a directional coupler to separate the incident and reflected signals (fig. 2(a)). The reflected signal is detected using an I-Q mixer, which enables the in-phase (I) and quadrature (Q) components of the reflected microwave signal to be detected with respect to the mixer's local oscillator (LO). The microwave frequency is adjusted to bring the I-component exactly in-phase with the local oscillator. Any $V_{\omega}(t)$ generated by the sample will be superposed on the much larger reflected signal. In order to determine the contribution of $V_{\omega}(t)$ we perform a lock-in experiment, low-frequency ($\sim$~44 kHz) pulse modulating the current through the sample and detecting the mixer outputs with a pair of lock-in amplifiers. Finally we extract the I ($V_{\omega,I}$) and Q components of the microwave voltage at the sample ($V_{\omega,Q}$).

The resulting signals are shown in figure 3(a). Since the sample is $\lambda/2$ away from the coupling capacitor the microwave current in the sample is nearly in phase with the reflected microwave signal. This means that the I channel has the form of $\chi_{yy}$, giving an anti-symmetric Lorentzian similar to the rectification measurement. Correspondingly the Q-component gives follows the form of $\chi_{yy}'$ and can be fitted to a Lorentzian. By dividing the maximum of $V_{\omega,Q}$ by $I_0$ we find the amplitude of the oscillating resistance during magnetisation precession to be $\delta R=4~m\Omega$. Using the in-plane AMR, measured at 1.1 \% for a similar sample at 30 K ($\Delta R = 100~\Omega$), we deduce an in-plane cone angle ($\approx \sqrt {\delta R/\Delta R}$) of 0.4 degrees. As the magnitude of $I_0$ is increased so does the amplitude of $V_\omega(t)$ (fig. 3(b)), from equation 2 we expect that $V_{\omega,Q(I)}\propto I_0$ and this is indeed observed (fig. 3(c)). The amplitude of $V_\omega$ should also be proportional to $B_1$, and in our case $B_1$ is a current induced effective magnetic field proportional to $I_1$. Hence we expect, and observe, that $V_{\omega,Q(I)}\propto I_1$ (fig. 3(d)).

We described how the capacitatively coupled $\lambda/2$ microstrip resonator enables the extraction of the microwave voltage generated by magnetisation precession in high resistance samples. This simple impedance transformer could also be applied to quantum circuits where the microwave conductance is of interest \cite{Gabelli_2006} and used as alternative to other transmission line matching techniques \cite{Hellmuller_2012,Hellmann:2012_a}.

The authors would like to thank T. Jungwirth, J. Wunderlich, H. Huebl and G. Puebla-Hellmann for comments on this manuscript. A.J.F acknowledges support from the Hitachi research fellowship and a Royal society research grant (RG110616).

%

\end{document}